\documentclass[prb,preprint,amsfonts,amssymb,amsmath,showpacs]{revtex4}

\usepackage{graphicx}
\usepackage{subfigure}
\usepackage{dcolumn}
\usepackage{bm}

\newcommand{\ket}[1]{\ensuremath{\left|#1\right>}}
\newcommand{\bra}[1]{\ensuremath{\left<#1\right|}}
\newcommand{\up}{\uparrow}
\newcommand{\down}{\downarrow}
\newcommand{\imag}{\mathtt{I\!m}}
\newcommand{\real}{\mathtt{R\!e}}

\begin{document}

\title{Magnetism in the Hubbard model-An improved treatment}

\author{A.Uldry}
\email{a.uldry1@physics.ox.ac.uk}
\author{R.J.Elliott}
\email{r.elliott1@physics.ox.ac.uk}
\affiliation{Department of Physics, Theoretical Physics, 1 Keble Road, Oxford OX1 3NP, England}

\date{\today}

\begin{abstract}
We propose a new treatment of the Hubbard model that is based both on the coherent-potential approximation (CPA) and the virtual-crystal approximation (VCA).
It is well known that the equilibrium found using the one-particle CPA 
Green's functions does not predict an ordered magnetic 
ground state, while Stoner's mean-field 
treatment, which is equivalent to the VCA 
on the Hubbard model, does so for a wide range of parameters. A hybrid 
treatment, the $\tau$-CPA, is developed, in which a particle is assumed to be 
scattered from an array of static opposite spins for a time $\tau$ related to 
the inverse of the band width. The propagation is treated in the CPA over 
this period; thereafter the particle sees the time-averaged effect of 
the scatterers and hence can be treated in the VCA. This model, with 
suitable approximations, does predict magnetism for a modified Stoner 
criterion. 
\end{abstract}


\maketitle

\section{\label{sec:intro}Introduction}
While much progress has been made in recent years in the theoretical treatment of magnetism 
in the transition metals and similar materials, a complete treatment of the many aspects of 
these familiar but complicated systems is still lacking.  The electrons retain many of their 
energy band properties as is reflected by their metallic behaviour, but the magnetic properties 
reveal that strong correlations exist between the electrons on the individual atoms, and exchange 
occurs between them. 

The simplest treatment of metallic ferromagnets is due to Stoner \cite{stoner} who assumed that 
the electron interaction gave rise to a molecular field which predicted ferromagnetism in a 
simple band model 
if the ratio of the interaction to the band width (or more precisely the density of states of the 
Fermi surface) was adequately large.  But beyond the mean field the model gives no correlation 
between the magnetic electrons. More complex band structures also predict antiferromagnetism in 
some cases \cite{penn}. 

In recent years much more sophisticated treatments of the band structure using established 
methods such as those based on the spin dependent local density 
approximation \cite{lsda_early,lsda_sic,lsda_U,lda_pp}
 have been able to make much 
better predictions of the Fermi surface and related magnetic properties, but again a full 
treatment of the atomic electron correlations is lacking.  Hybrid treatments such as that of 
Gutzwiller \cite{gutzori,gutznew,gutzmulti} which start from a mixture of correlated and uncorrelated states can give an 
improved treatment of the magnetic properties at the expense of some of the metallic ones.  

An alternative approach to this problem has been through the study of simple band models with electron interactions of 
which that due to Hubbard \cite{hubbI} is the simplest which appears to contain the essence of the problem.  
He restricts the electron-electron interaction to occur on single atomic sites which in the case 
of a single band only occurs between electrons of opposite spin because of the exclusion principle.  
Hubbard, in a series of papers \cite{hubbI,hubbIII}, pioneered different approximate treatments to this model Hamiltonian. 
The simplest replaces the electron-electron interaction by its average so that an electron of one spin 
sees a field proportional to the density of the other spin and the system has simple energy bands 
separated by an energy proportional to the magnetisation and hence is completely equivalent to 
Stoner's mean field approximation.  In a different vocabulary it can be regarded as a virtual crystal 
approximation (VCA) where the electrons move in energy bands derived from an average potential.  

In order to give a better treatment of the electron-electron scattering 
Hubbard \cite{hubbIII} pioneered another 
approximation where it was assumed that electrons of one spin were frozen on their sites and the 
electrons of the other spin moved through the random potential so created.  He proceeded to solve 
this problem by determining an effective potential (which was complex and energy dependent) which 
took account of the average scattering from this array in an analogy with an alloy.  This method has 
been independently used in a wide variety of alloy problems and is usually referred to as the 
coherent potential approximation (CPA) \cite{lax,taylor,soven,elliott}.  It is clear that this approximation over-estimates the 
electron-electron scattering and in particular causes the energy bands of each spin character to 
split if the interaction is large enough.  As a result the tendency to ferromagnetism is 
under-estimated and it has been demonstrated \cite{schneidrchal,harrislange} that no magnetic order is predicted except in 
the special case of a half filled band.

Several attempts have been made to improve on these approximate treatments, for example using a 
dynamical CPA \cite{kake92,kake02} based on the integration of this approximation into functional integral 
techniques.  There has been a number of other attempts to modify the alloy analogy \cite{sda,maa,maaeha,maacond}.  
In this work we propose a simple physical model which is intermediate between the straightforward 
VCA and CPA treatments.  In it we assume that the random scattering centres experienced by an 
electron of one type in the alloy analogy exist only for a certain time $\tau$ which is proportional 
to a typical hopping time between sites and hence inversely to the band width.  At later times the 
electrons see a completely average potential reflecting the movement of the scattering centres.  
Thus the propagators are calculated in the CPA approximation for the short time interval $\tau$ and have 
to be matched to the VCA propagators which hold for longer times.  Although this is a 
straightforward procedure in principle certain further approximations are necessary to 
make it tractable.  The properties of the system are evaluated in this so-called $\tau$-CPA.  
The single particle properties such as the density of states show,
as expected, a mixture of characteristics associated with the VCA and CPA.  
Ferromagnetism is predicted to occur for a range of parameters which are somewhat 
more restrictive than those in the simple Stoner model. \\
This paper is organised as follows: Sec.\ref{sec:model} introduces the model used. The VCA and CPA methods are outlined in Sec.\ref{sec:vcacpa}. The $\tau$-CPA, mixing the VCA and CPA, is constructed in Sec.\ref{sec:taucpa}. Numerical results for the density of states and the magnetisation using a simple cubic band are presented and discussed in Sec.\ref{sec:results}.

\section{\label{sec:model}Model Hamiltonian}
The Hubbard model is currently widely considered to be the most concise model that captures the essence of the interactions in an electronic system. It allows the hopping of the electrons from site to site to compete with the Coulomb repulsion while including the Pauli principle. We will consider the simplest case of the Hubbard model, a single-band with nearest-neighbour hopping $t_{ij}$ and on-site repulsion :
\begin{equation} \label{hubbard}
H = \underbrace{\sum_{i,j,\sigma}
t_{ij}c^{+}_{i\sigma}c_{j\sigma}}_{H_{0}}
+ U\sum_{i}c^{+}_{i\uparrow}
c_{i\uparrow}
c^{+}_{i\downarrow}c_{i\downarrow}
\end{equation}
$c^{+}_{i\sigma}$ and $c_{i\sigma}$ are the usual creation and destruction operators for electrons in the Wannier states. The free part of the Hamiltonian $H_{0}$ is evaluated in the Bloch states $H_{0}=\sum_{k}\epsilon_k c^{+}_{k\sigma}c_{k\sigma}$ using a tight-binding approximation for the band structure $\epsilon_k$. We will concentrate in this paper on the simple cubic band of half band width $w=1$. As for the interaction, it is repulsive and only felt when one electron of spin $\sigma$ meets another of
spin $-\sigma$ at the same site. This model simulates relatively well a real
material possessing narrow energy bands, like the transition metals
and their alloys. We will work on a  three dimensional lattice of $N=L^3$ sites, at $T=0$. We use the framework of the canonical ensemble and fix the 
number $N_e$ of electrons per atom that distribute themselves
at random over the lattice sites. The Fermi statistics takes care that
no site is occupied by two electrons of the same spin while the
interaction favours singly occupied sites. The propagation of an electron of spin $\sigma$ in the lattice can be described by the retarded Green's function
\begin{align}\label{tGFdef}
G^{\sigma}_{ij}(t,t') &=-i\,2\pi \, \theta(t-t')\langle \{ c_{i\sigma}(t),
c_{j\sigma}^{+}(t')\}_+ \rangle \nonumber \\
 &=:\langle \langle  c_{i\sigma}(t); c_{j\sigma}^{+}(t') \rangle  \rangle 
\end{align}
where $\hbar$ has been set to $1$. The following Fourier transform (where no particular origin is specified)
\begin{equation}\label{EGFdef}
G^{\sigma}(E)=\frac{1}{2\pi}\int_{-\infty}^{\infty}d(t-t')\,G^{\sigma}
(t-t')e^{iE(t-t')}
\end{equation}
will often be used, and $t'$ will be set to $0$. \\
We are interested in the properties of this system averaged over all possible
configurations, for which the Green's function $G_{k}^{\sigma}(E)$ can be 
defined by
\begin{equation}\label{kGFdef}
G^{\sigma}_{ij}(E)=\frac{1}{N}\sum_{k}G^{\sigma}_{k}(E)e^{ik(r_i-r_j)}
\end{equation}
The density of states $\rho ^{\sigma}$ is obtained from the single-particle 
Green's function with
\begin{equation}\label{dosrho}
\rho^{\sigma}(E)=\frac{-1}{\pi}\imag{F^{\sigma}(E)}
\end{equation} 
where $F^{\sigma}(E)$ is the trace of the Green's function. It can be obtained as
\begin{equation}\label{Fdef}
F^{\sigma}(E)=\frac{1}{N}\sum_k G^{\sigma}_k(E)
\end{equation}
The relative proportions of the number of up-spins and down-spins
must be determined self-consistently together with the chemical potential
$\mu$. Defining $N^{\up}$ and $N^{\down}$ respectively as the average number of spins up and down per atom, the equilibrium conditions read
\begin{equation}\label{nostates}
N^{\sigma}=\int_{-\infty}^{\mu}dE\,\rho^{\sigma}(E)
\end{equation}
with $\sigma \in \{\up,\down \}$

\section{\label{sec:vcacpa}VCA and CPA}
The virtual-crystal approximation and the coherent-potential approximation have been both applied with various success to both the Hubbard model and the somewhat related problem of the random binary alloy $A_{1-c} B_c$. It is worth recalling briefly the development of the VCA and the CPA in the binary alloy before turning to the Hubbard model.
\subsection{\label{subsec:alloy}VCA and CPA in the binary alloy}
The analogy between the Hubbard model and a binary alloy was noted by Hubbard in the last paper of the series \cite{hubbIII}. We consider a particle 
propagating on a lattice where each site is occupied at random either by a host atom $A$ or an impurity atom $B$. The concentration of 
impurities is $c$. The model is represented by $H_{alloy}=H_o+V$, where $V$ is a quadratic operator of the form $V=\sum_{m} \ket{m}\epsilon_m \bra{m}$. The atomic potentials can take two values, either $\epsilon_A$ or $\epsilon_B$, depending on whether the particle is on an $A$ site or a $B$ site. Setting $\epsilon_A$ to $0$ and $\epsilon_B$ to $U$, we find that this is very similar to the Hubbard model (hence the alloy analogy), for which the spin $\sigma$ propagating in the lattice interacts with energy $U$ only on sites where a $-\sigma$ spin is present. The analogy would be exact if the spins of opposite direction could be considered as frozen in the Hubbard model.\\
The VCA and CPA are well established methods \cite{velickirk,elliott} and will only be outlined here.\\
If $P$ describes the propagator in the perfect crystal defined by $H_0$, the perturbation expansion in the whole Hamiltonian $H_{alloy}$ reads
\begin{equation}\label{pertexp}
G =P+PVP+P\,VPVP+\cdots =P+PVG
\end{equation}
The problem of evaluating the configurational average $\langle G \rangle= P+P \langle VG \rangle$ is solved in the VCA by the random phase approximation decoupling $\langle V G\rangle \approx \langle V \rangle\langle G \rangle$. With $\langle \epsilon_m\rangle = cU$, the propagator for the binary alloy in the VCA is
\begin{equation}\label{VCAalloy}
G_{k}^{VCA}(E)=\frac{1}{E-\epsilon_k-cU}
\end{equation}
While the VCA is only a valid approximation for small perturbations $V$, 
the CPA is a particularly successful method that interpolates between the limits of strong and weak disorder and interaction. The CPA assumes the existence of an effective 
medium described by the Green's function $G_e$ and self-energy $\Sigma$ that obey the Dyson equation
\begin{equation}\label{GeDyson}
G_e=P+P\Sigma G_e
\end{equation}
The true Green's function $G$ can be developed in the perturbation expansion 
(\ref{pertexp}), which, together with (\ref{GeDyson}), lead to a new expansion of $G$ in terms of the medium
\begin{equation}
G=G_e+G_e\left(V-\Sigma \right)G
\end{equation}
A particle moving in the effective medium sees the impurities as embedded in a uniform potential. As shown in Refs.\onlinecite{velickirk,elliott}, it is also possible to expand the true $G$ in terms of $G_e$ and a scattering $T$ matrix so that
\begin{equation}\label{Tmat}
G=G_e+G_e TG_e
\end{equation}
The $T$ matrix can be expressed as a function of the individual atomic scattering matrices $T_m$ 
\begin{equation}\label{TintoTm}
T=\sum_m T_m + \sum_{n\neq m}T_n G_e T_m + \sum_{n\neq m \neq l}T_n G_e T_m G_e T_l +\cdots
\end{equation}
Projected on the the Wannier states, the scattering on the $m$ atom can be found to be \cite{velickirk,elliott}
\begin{equation}\label{cpatm}
t_{m}=\frac{\epsilon_m-\Sigma}{1-(\epsilon_m-\Sigma)F}
\end{equation}
$F$ is the diagonal element of $G_e$, $F=\bra{m}G_e\ket{m}$ (see \ref{Fdef}). A central step of the CPA is to demand that the medium propagator $G_e$ corresponds to the true propagator averaged over the disorder: 
$G^{CPA}:=\langle G \rangle \equiv G_e$. Consequently, from (\ref{Tmat}), the total scattering of impurities is set to zero: $\langle T\rangle=0$. $\langle T \rangle$ can then be decoupled by 
a random phase approximation on the single $T_m$ matrices. We note that this is a much better approximation than the VCA, since the repeated scattering by the same site is forbidden, as (\ref{TintoTm}) shows. In this form, the CPA neglects the scattering from \textit{clusters} of atoms, though some extensions of the theory do include this effect. \\
The CPA condition becomes therefore
\begin{equation}\label{cpatmcond}
\langle t_m \rangle=0
\end{equation}
This is the central equation of the CPA.
The self-energy is calculated such that the scattering generated by any single impurity is zero on average. The CPA approximation for the one particle Green's function is then given by
\begin{equation}\label{onepartCPA}
G_k^{CPA}(E)=\frac{1}{E-\epsilon_k-\Sigma(E)}
\end{equation}
with the self-energy $\Sigma$ determined self-consistently with $F$ from (\ref{cpatmcond}) and (\ref{cpatm}). From these two equations and for the binary allow with $\epsilon_A=0$, 
$\epsilon_B=U$, the relations between $F$ and $\Sigma$ can be expressed as
\begin{equation}\label{sigcpa}
\Sigma=\frac{cU}{1-(U-\Sigma)F}
\end{equation} 
where $F$ is the diagonal element of $G_k^{CPA}$.
While the VCA self-energy is a real number that merely shifts the energy levels by $cU$, the CPA 
$\Sigma$ is an energy-dependant, complex number which accounts for the damping of 
the quasiparticle states by the impurities. 

\subsection{\label{subsec:vcacpahubb}VCA and CPA in the Hubbard model}
The VCA for the Green's function (\ref{tGFdef}) in the Hubbard model (\ref{hubbard}) can be obtained in two different ways. The first approach consists in applying the Hartree-Fock approximation to 
the Hubbard model and effectively reducing it to a Stoner model
\begin{equation}\label{stonermod}
H_{Stoner} =H_0+U\sum_{i}\left(N^{\up} \hat{n}_{i\down}+ 
N^{\down} \hat{n}_{i\up}  \right)
\end{equation}
The usual number operator notation $\hat{n}_{i\sigma}=c^{+}_{i\sigma}c_{i\sigma}$ has been used. The Stoner model (\ref{stonermod}) leads directly to the set of Green's function analogous to (\ref{VCAalloy})
\begin{equation}\label{vcaGF}
G^{\sigma \,VCA}_k(E)=\frac{1}{E-\epsilon_k-N^{-\sigma}U}
\end{equation} 
The up-spins see the (frozen) down-spins as impurities of concentration $N^{\down}$ with whom they interact with energy $U$, and vice-versa for the down-spins.\\ 
The second way of obtaining the virtual-crystal approximation is to decouple the equation of motion for the Green's function (\ref{tGFdef}) in the Hubbard model
\begin{align}\label{teom}
i\frac{d}{dt}\langle \langle c_{i\sigma}(t), c^{+}_{j\sigma} 
\rangle \rangle &=2\pi \delta (t)\delta_{ij}+\sum_{l}t_{il}\langle \langle c_{l\sigma}(t), c^{+}_{j\sigma}\rangle \rangle \nonumber \\ 
 &+ U\langle \langle \hat{n}_{i,-\sigma}(t)c_{i\sigma}(t), 
c^{+}_{j\sigma} \rangle \rangle
\end{align}
If $\hat{n}_{i,-\sigma}(t)$ is taken out of the last Green's function in 
(\ref{teom}) and averaged separately, $\langle \hat{n}_{i,-\sigma}(t) \rangle \rightarrow N^{-\sigma}$, the equation of motion is decoupled and the solution is given again by (\ref{vcaGF}). If, on the other hand, $ \hat{n}_{i,-\sigma}(t)$ is considered at $t=0$ and takes the value $0$ or $1$ at random, applying thus the alloy analogy, the CPA result (\ref{onepartCPA},\ref{sigcpa}) can be retrieved. 
\begin{gather}\label{cpaGF}
G^{\sigma \,CPA}_k(E)=\frac{1}{E-\epsilon_k-\Sigma^{\sigma}(E)}\nonumber \\
\Sigma^{\sigma}(E)=\frac{N^{-\sigma}U}{1-[ U-\Sigma^{\sigma}(E)] F^{\sigma}(E)}
\end{gather}
This is however not a straightforward procedure. Hubbard developed an elaborate scheme based on the alloy analogy in the last paper of the series \cite{hubbIII}, calling it the ``scattering correction''. It was noted later by Velick\'y \textit{et al} that the scattering correction Green's function is the same as the CPA result for the alloy model. 

\section{\label{sec:taucpa}The $\tau$-CPA}
It has been known for some time now that the CPA does not allow for ferromagnetic solutions \cite{harrislange,schneidrchal}. On the other hand, the VCA treatment of the Hubbard model leads to an artificially strong 
ferromagnetic phase. We propose to combine these two different 
approximations in order to keep the advantages of the CPA while allowing for spontaneous magnetic ordering to occur. As was seen in the previous section, the VCA and CPA take two very different views when considering the motion of a particle in the lattice. The CPA sees the scatterers as fixed at the positions given at $t=0$, while the VCA considers the scattering events only as averaged over a time $t\rightarrow \infty$. In other words, the CPA is valid for the Hubbard model only at $t=0$ and the VCA at $t\rightarrow \infty$. A better physical picture should emerge if the CPA treatment is applied for a short time only, when it is still reasonable to consider the scatterers as fixed. This time is typically the time needed for the particle to travel between sites. Thereafter we can assume that the VCA average provides a reasonably good treatment. \\

Let $G_k^{VCA}(t)$ and $G_k^{CPA}(t)$ be respectively the time Fourier transform of the VCA and CPA Green's functions for $t\geq0$. The central assumption is that up to a time $\tau > 0$ and $\tau \approx 1/w$ the system is described by the CPA Green's function, whereas beyond  $\tau$ it is described by the VCA Green's function. A new function  $g_k^{\tau CPA}(t)$ is defined as
\begin{equation} \label{taucpamod}      
g_k^{\tau CPA}(t)=-i 2\pi\, \theta (\tau -t) G_k^{CPA}(t)-\alpha \,i 2\pi \, \theta (t-\tau) G_k^{VCA}(t)
\end{equation}
At $t=0$, $g_k^{\tau CPA}$ is equal to the CPA Green's function and is thus properly normalised. A factor $\alpha$, the ``matching factor'', has been introduced to ensure the continuity of the wave function at time $\tau$. \\
The retarded Green's functions for the VCA and CPA are given respectively by $G_k^{VCA}(E)$ 
(\ref{VCAalloy}) and $G_k^{CPA}(E)$ (\ref{onepartCPA}). The Fourier transform reads:
\begin{equation} \label{gtimeFT}      
G_k(t) = \int_{-\infty}^{\infty}dE e^{-iE t}G_k(E)
\end{equation}
It is straight-forward to obtain $G_k^{VCA}(t)$ by contour integration in the lower half of the complex plane. 
There is one pole at $z_0=\epsilon_k+cU-i\delta$, so that
\begin{equation} \label{gtvca}      
G_k^{VCA}(t)=-i 2\pi \,\theta (t)e^{-i(\epsilon_k+ c U)t}
\end{equation}
It is more difficult to do a similar calculation for the CPA. The problem lies in the nature of the CPA poles. It has been noted by Velick\'y \textit{et al} 
\cite{velickirk} that the two complex CPA poles cannot in all generality be interpreted as quasiparticle energy. A calculation for the CPA similar to the one made for the VCA does not lead to a correct result. In particular, the correct inverse Fourier transform is not retrieved. In order to keep the denominator of the energy Green's function as $E-\epsilon_k-\Sigma(E)$, we assume instead, by similarity with the VCA:
\begin{equation} \label{gtcpa}      
G_k^{CPA}(t)=-i2\pi\, \theta (t)e^{-i(\epsilon_k+ \Sigma)t}
\end{equation} 
At this stage the fact that $\Sigma$ is an energy-dependant quantity is ignored, and $\Sigma$ is treated as a fixed quantity rather than a function. By doing so, Fourier-transforming (\ref{gtcpa}) gives the expected result (\ref{onepartCPA}). The substitution $\Sigma \rightarrow \Sigma(E)$ must be made thereafter. The whole $g^{\tau CPA}_{k}(t)$ is still normalised at $t=0$, but because of the dependence of $\Sigma$ on the energy, this does not guarantee the normalisation of $g^{\tau CPA}_{k}(E)$. The normalisation will have to be corrected 
numerically. \\
The integrals for the Fourier transform on the partial intervals give, without taking into account the matching factor $\alpha$,
\begin{gather}
\mathcal{G}_{k}^{VCA}(E)=\frac{1}{2\pi}\int_{0}^{\tau}dt \,e^{iEt}G_{k}^{VCA}(t)=
\frac{e^{i\left(E-\epsilon_k-cU  \right) \tau}}{E-\epsilon_k-cU } \label{VCAnat} \\
\mathcal{G}_{k}^{CPA}(E)=\frac{1}{2\pi}\int_{\tau}^{\infty}dt \,e^{iEt}G_{k}^{CPA}(t)=
\frac{1-e^{i\left(E-\epsilon_k-\Sigma  \right) \tau}}{E-\epsilon_k-\Sigma } \label{CPAnat}
\end{gather}
The $\mathcal{G}_{k}^{VCA}(E)$ and  $\mathcal{G}_{k}^{CPA}(E)$ are the standard VCA and CPA Green's functions, weighted by a complex $\tau$-dependant quantity.\\
The factor $\alpha$ that forces the CPA and VCA Green's functions to match at $t=\tau$ is found to be
\begin{equation} \label{alpha}      
\alpha =e^{-i(\Sigma -c U)\tau}
\end{equation}
This value for $\alpha$ can be reintroduced into (\ref{taucpamod}). The result for $g^{\tau CPA}_{k}(E)$ is then obtained from the Fourier transform of (\ref{taucpamod}):
\begin{align} \label{gtaunonorm}      
g_k^{\tau CPA}(E) &= \frac{1}{2\pi}\int_{0}^{\infty}dt \,G_k^{\tau CPA}(t)e^{iE t} \nonumber\\
 &= \frac{1-e^{i(E-\epsilon_k-\Sigma)\tau}}{E-\epsilon_k-\Sigma}+
\frac{e^{i(E -\epsilon_k-\Sigma)\tau}}{E -\epsilon_k-c U}
\end{align}
$\Sigma$ is now read as $\Sigma(E)$, as calculated by the pure CPA equation 
(\ref{sigcpa}). 
The expression (\ref{gtcpa}) now reproduces exactly the CPA behaviour in the limit $\tau \rightarrow \infty$. The VCA approximation is retrieved in the limit 
$\tau \rightarrow 0$. \\

We recall that the normalisation of the Green's function has been lost by choosing to ignore the energy dependence of $\Sigma$. It is therefore 
necessary to calculate numerically for each $k$ the factor $\beta_k$, so that
$\beta^{-1}_k g_k^{\tau CPA}(E)$ is normalised. \\
The result (\ref{gtaunonorm}) is however not giving a physical density of states: $\rho (E)$ changes sign along the $E$ axis. 
By stopping the Green's function at a time 
$\tau$ we have in fact introduced a non-physical oscillatory term coming from the factor weighing VCA and CPA 
\begin{equation} \label{oscill}
e^{i(\omega -\epsilon_k-\Sigma)\tau}=e^{i(\omega -\epsilon_k-\real{\Sigma})\tau} e^{\imag{\Sigma}\tau}
\end{equation}
The oscillatory term destroys the analytical properties of the CPA and VCA. 
The imaginary part of $N^{-1}\sum_{k}-\pi^{-1} g^{\tau CPA}_k(E+i\delta)$ does not give the physical density of states, as the 
branch cut no longer runs along the x-axis. It is thus necessary to smooth out the Green's function by putting the oscillatory term to unity. We obtain finally 
\begin{equation} \label{gwtau}      
G_k^{\tau CPA}(E) = 
\frac{1}{\beta_k}\left( \frac{1-e^{\imag{\Sigma}(E)\tau}}{E -\epsilon_k-\Sigma(E)}+
\frac{e^{\imag{\Sigma}(E)\tau}}{E -\epsilon_k-c U}\right )
\end{equation}
This function now possesses all the properties required. It interpolates between a pure VCA Green's function at $\tau \rightarrow 0$ and the CPA Green's function at $\tau \rightarrow \infty$. We observe that (\ref{gwtau}) is decomposed into a CPA and VCA contribution weighted by a real $\tau$-dependant factor. Although $\tau$ is not strictly determined by the problem, a reasonable value for it is given on physical ground by the band width.

\section{\label{sec:results}Numerical results and discussion}
In this section, the $\tau$-CPA Green's function (\ref{gwtau}) is used to obtain the density of states for both species of spins in the Hubbard model. Various values of the impurity concentration $c$ ($c$ corresponds to either $N^{\up}$ or $N^{\down}$), of the interaction $U$ and of $\tau$ are considered. Calculations are made for a simple cubic band with $w =1$. Another one-particle property is then evaluated, the magnetisation $m$ versus the interaction strength $U$, for various range of parameters. 

\subsection{\label{subsec:dos}Density of states in the $\tau$-CPA}
\begin{figure}
\includegraphics[width=8.6cm]{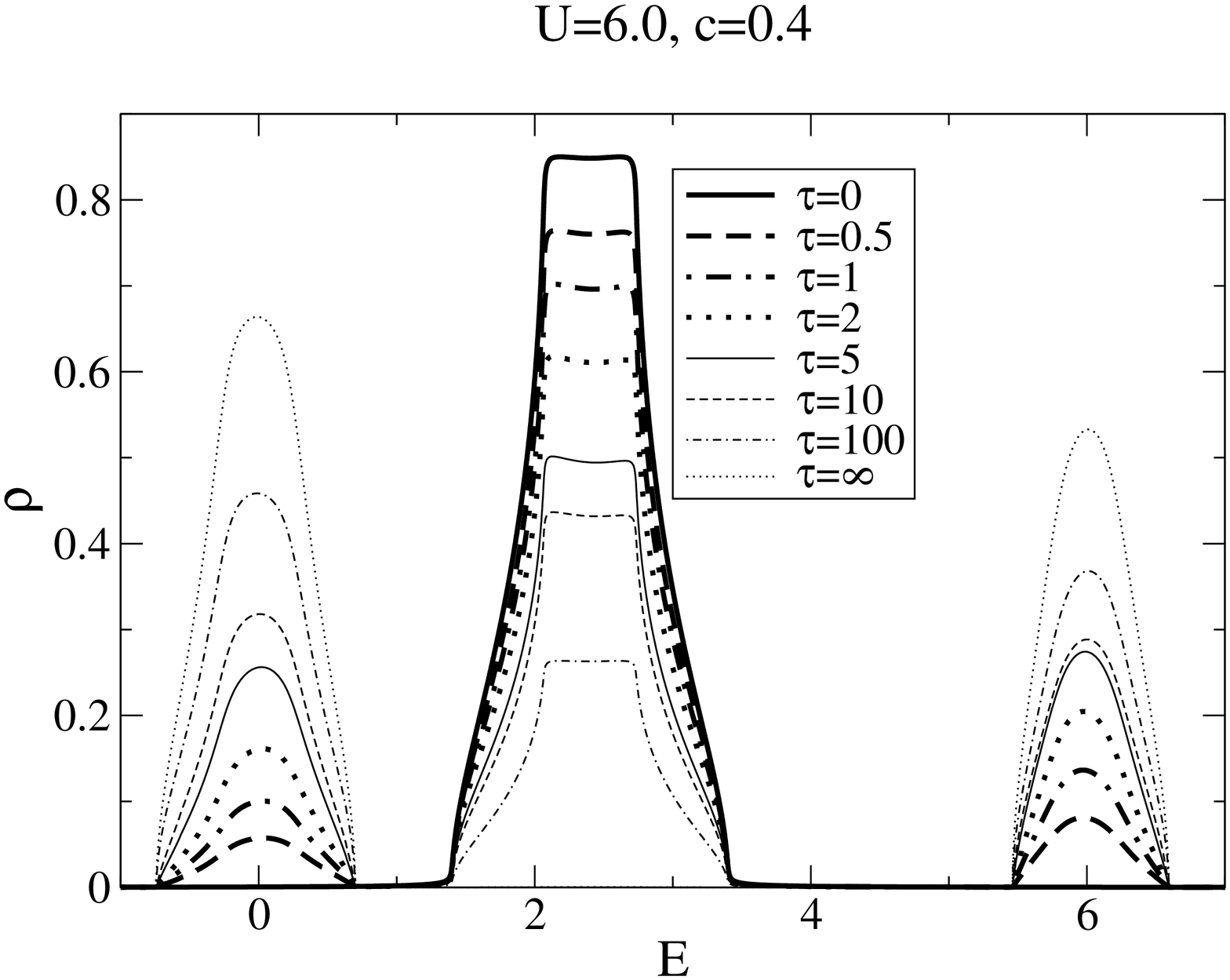}
\caption{DOS $\rho$ for the $\tau$-CPA using a simple cubic band for $U=6$, $c=0.4$ and $w=1$.\label{fig:tauU6dos}}
\end{figure}

The density of states is obtained from the Green's function (\ref{gwtau}) using (\ref{dosrho},\ref{Fdef}). Fig.\ref{fig:tauU6dos} shows the density of states obtained for a strong disorder and interaction strength, $U=6.0$ and $c=0.4$. 
 \begin{figure}
\includegraphics[width=8.6cm]{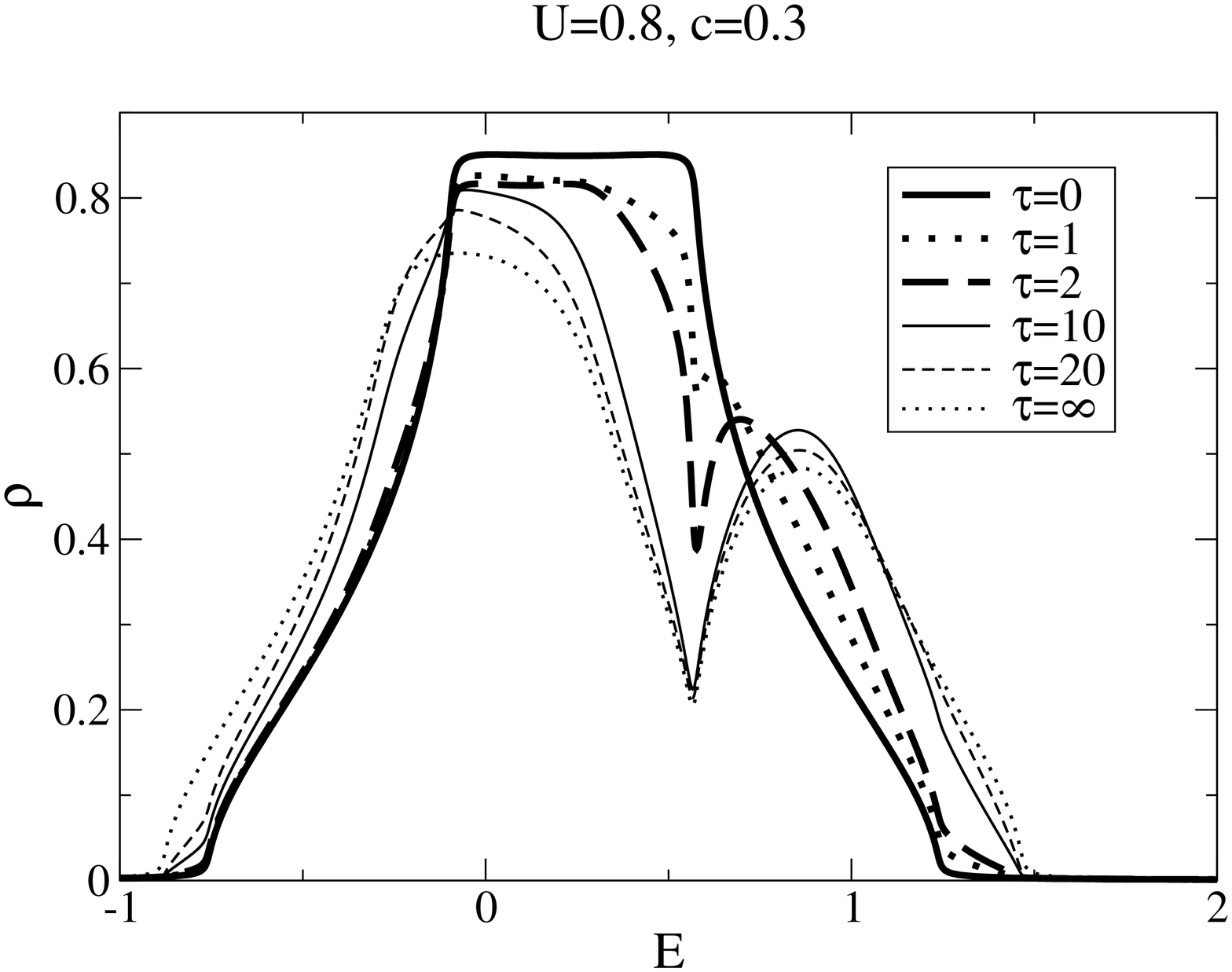}
\caption{DOS $\rho$ for the $\tau$-CPA using a simple cubic band for $U=0.8$, $c=0.3$ and $w=1$.\label{fig:tauU08dos}}
\end{figure} 
\begin{figure*}
\subfigure[small values of $\tau$]{\label{fig:tauU2dos1}
\includegraphics[width=8.6cm]{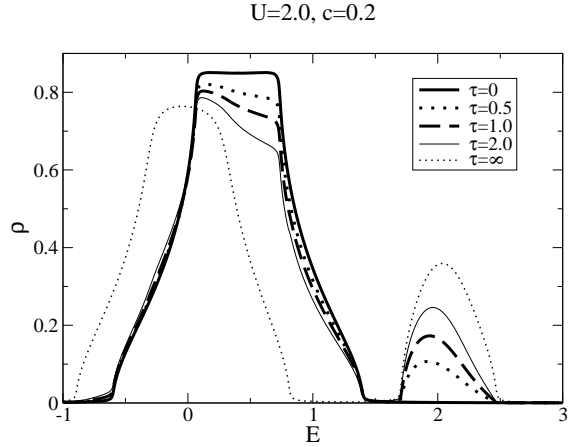}}
\subfigure[large values of $\tau$]{\label{fig:tauU2dos2}
\includegraphics[width=8.6cm]{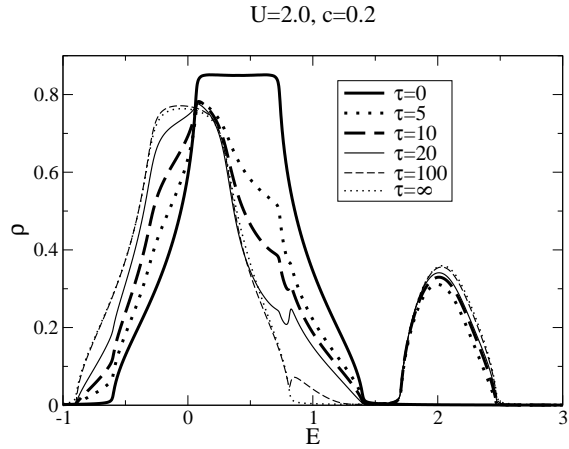}}
\caption{DOS $\rho$ for the $\tau$-CPA using a simple cubic band for $U=2$, $c=0.2$ and $w=1$.\label{fig:tauU2dos}}
\end{figure*}
$\tau$ is varied from $0$ (pure VCA) to $\tau \rightarrow \infty$ (pure CPA).
At $\tau=0$ the spectrum exhibits the single VCA band, centred at $E=cU$. As $\tau$ increases, the two CPA sub-bands appear, at $E=0$ for the main band and 
$E=U$ for the impurity band. Weight is transfered from the VCA band to both the CPA sub-bands, affecting all the three sub-band heights, but not their general shape.
At $\tau \rightarrow \infty$ the VCA band has disappeared and only the two CPA sub-bands survive, with their original, pure CPA respective weight. In this graph the 
$\tau$-CPA merges with the pure CPA at $\tau \approx 600$. \\
More dramatic effects on the band shapes occur at smaller $U$, when the bands mix. Fig.\ref{fig:tauU08dos}, for $U=0.8$ and $c=0.3$, depicts a case for which $U$ and $c$ are sufficiently small so that the pure CPA sub-bands are not split. As $\tau$ increases, the height of the VCA band decreases while that of the main CPA 
sub-band increases. 
The general picture produced in that case is that for $\tau > 0$, the degenerate states near the centre of the VCA band start to scatter, while the edges of the band are still untouched. For larger $\tau$, the states continue to diffuse to the impurity band from the centre to the right edge. At the same time, states of lower energy are being occupied, re-centring the created sub-band around $0$.
Cases at intermediate $U$ are not so smooth, as can be seen in 
Figs.\ref{fig:tauU2dos1} and \ref{fig:tauU2dos2}, where $U=2.0$ and $c=0.2$. 
In this figure the density of states of the $\tau$-CPA becomes equal to that of the CPA for $\tau \approx 250$. The mixing of the bands produces irregular features as the diminishing VCA meets the increasing main CPA 
band at larger $\tau$, Fig.\ref{fig:tauU2dos2}. This is due to the fact that a large imaginary part at an energy $E$ favours the CPA behaviour compared to that of the VCA. The CPA self-energy has an imaginary part only within the CPA bands. When the VCA band is centred within the main CPA band but is sufficiently 
distinct from it, its flat top can be sharpened if close to the top of the CPA band, producing the small peaks in Fig.\ref{fig:tauU2dos2}. Such drops in the VCA part of the density of states also occur for the parameters of Fig.\ref{fig:tauU08dos}. However under these conditions the VCA band is only slightly shifted from the CPA band, and the sharp peaks are compensated by the CPA part.

\subsection{\label{subsec:magnetisation}Magnetisation in the $\tau$-CPA}
\begin{figure}
\includegraphics[width=8.6cm]{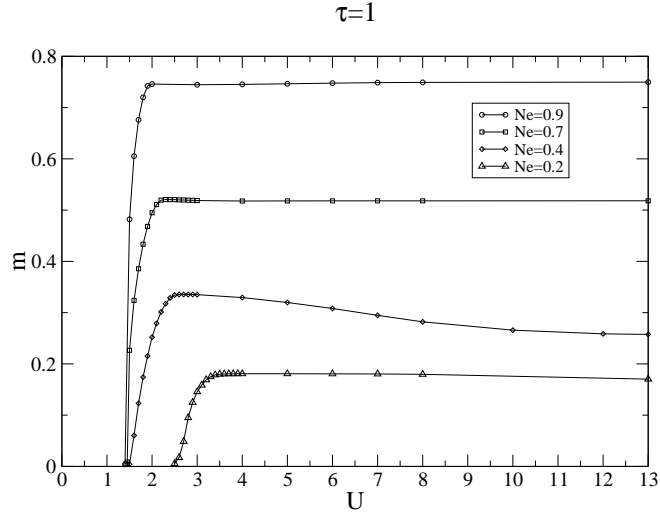}
\caption{Magnetisation $m$ versus interaction $U$ for the $\tau$CPA at $\tau=1$ and various filling $N_e$.\label{pict:mUtau1}}
\end{figure}

\begin{figure}
\includegraphics[width=8.6cm,height=10cm]{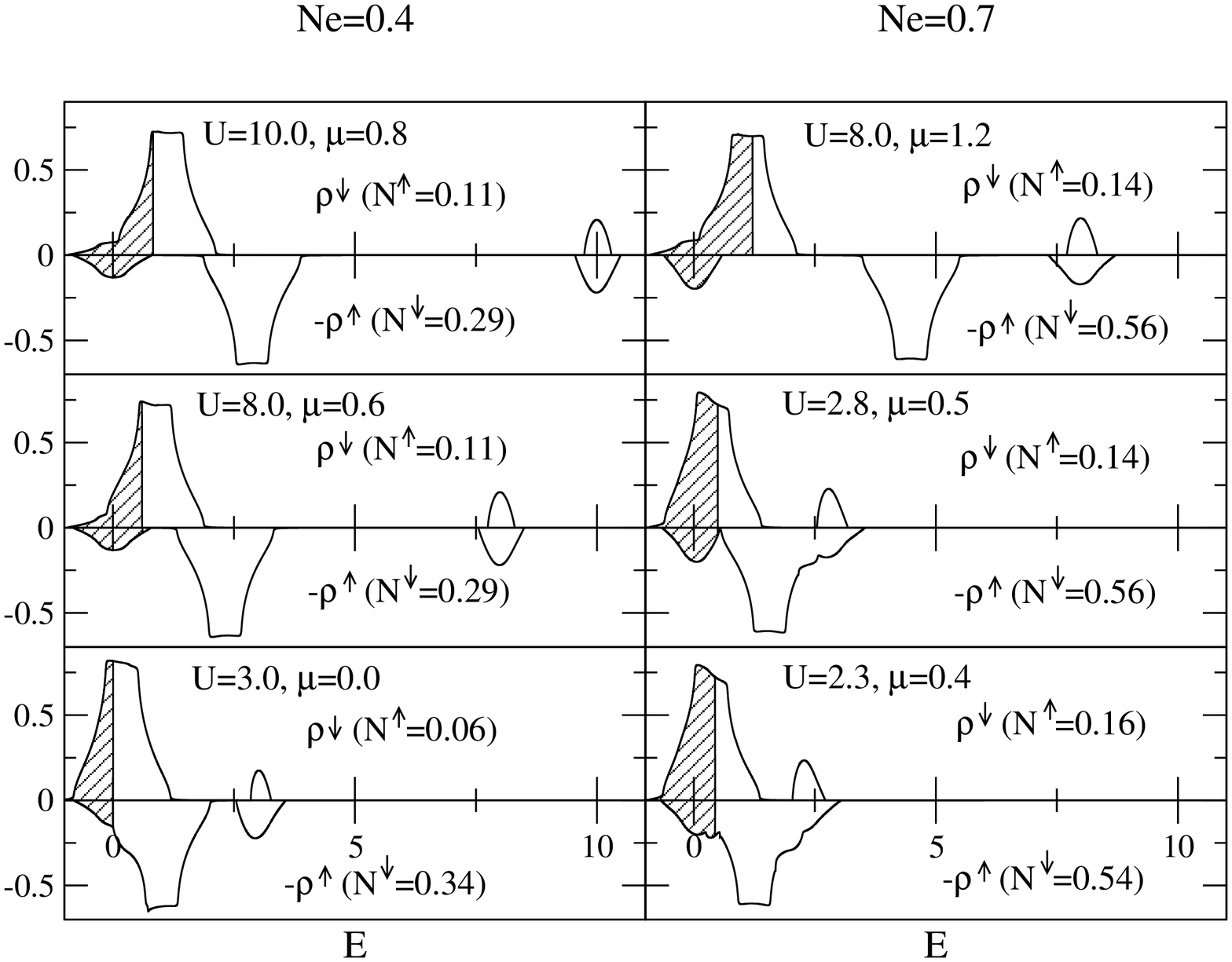}
\caption{Density of states for both the up and down particles in the $\tau$CPA with $\tau=2.0$. On the left the concentration is kept at $N_e=0.4$ for 
$U=10.0$, $8.0$ and $3.0$. On the right $N_e=0.7$ and $U=8.0$, $2.8$ and $2.3$.  Occupied states are marked by the hatched area.\label{fig:dostau2}}
\end{figure}

The $\tau$-CPA offers a treatment that leads to density of states that interpolate between the pure VCA and the pure CPA. As will be shown in this section, the method also interpolates the value of the magnetisation versus the interaction strength $U$ between the VCA ground state (ordered ground state for sufficiently large $U$) and the CPA ground state (no ordered ground state at any $U$). We only consider in this paper the competition between paramagnetic and ferromagnetic ground states. \\
 In practice, the equilibrium conditions for a certain $U$ and a given, fixed number of spins $N_e$, are found by iteration. For example, a value for  $N^{\up}$ is given as an input. This allows the determination of 
$G^{\down}$ (\ref{gwtau}) and the
density of states from (\ref{dosrho},\ref{Fdef}). $N^{\down}$, deduced from 
$N_e=N^{\up}+N^{\down}$, determines the chemical potential $\mu$ from 
(\ref{nostates}) with $\sigma=\down$. $N^{\down}$ also generates $G^{\up}$, which together
with the chemical potential, determine a new value for $N^{\up}$. The
procedure is repeated until $N^{\up}$ converges. 
The paramagnetic solution $N^{\up}=N^{\down}$ is always a solution. If
another solution is found, its stability must be tested against the
paramagnetic case by measuring the total energy of both solutions. 
The total energy $E$ of the system is found by summing up the separate 
energies of the two spins \cite{schneidrchal,sda}
\begin{equation}
E=\frac{1}{2}\int_{-\infty}^{\mu}d\omega \sum_{k,\sigma}
(\omega+\epsilon_k)\,A_{\sigma}(k,\omega)
\end{equation}
where $A_{\sigma}(k,\omega)$ is the spectral density defined by 
$A_{\sigma}(k,\omega)=-\pi^{-1}\imag{G^{\sigma}_{k}}(\omega)$.
Diagrams of magnetisation $m=N^{\down}-N^{\up}$ versus interaction strength $U$ can be calculated for any value of $N_e$.
In Fig.\ref{pict:mUtau1} the magnetisation is depicted for a range of values of $U$.
\begin{figure}
\includegraphics[width=8.6cm]{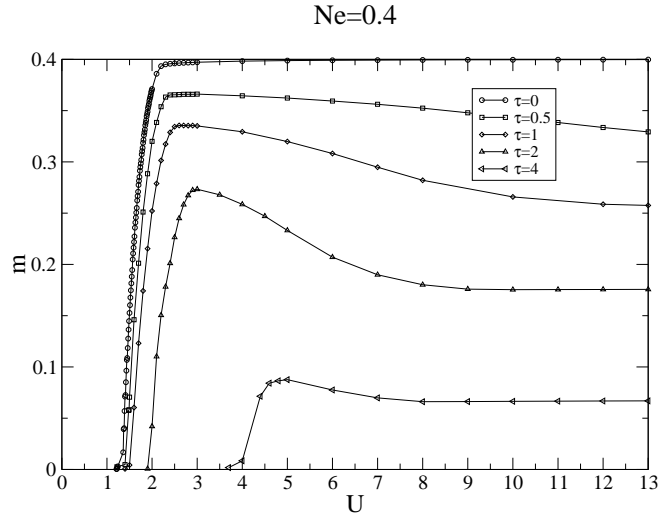}
\caption{Magnetisation $m$ versus interaction $U$ for the $\tau$CPA at various $\tau$ but a fixed filling of $N_e=0.4$. \label{pict:mUNe04}}
\end{figure}
\begin{figure}
\includegraphics[width=8.6cm]{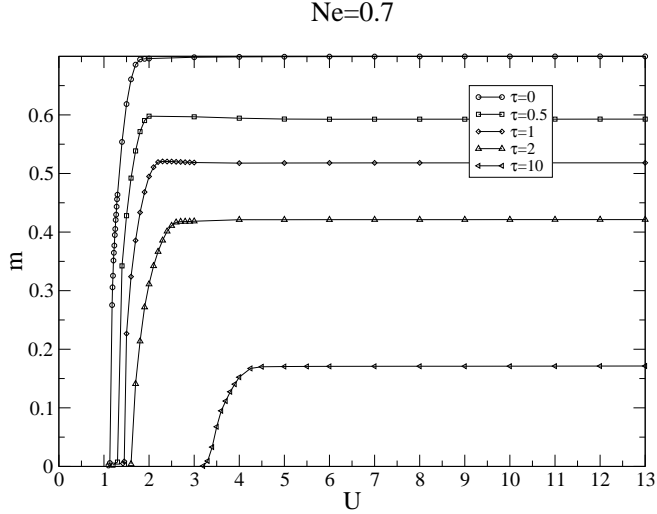}
\caption{Magnetisation $m$ versus interaction $U$ for the $\tau$CPA at various $\tau$ but a fixed filling of $N_e=0.7$. \label{pict:mUNe07}}
\end{figure}
 $\tau$ is fixed to $1$ and the curves calculated for various filling. The first observation is that at no value of $N_e$ ($N_e < 1$) is there saturated ferromagnetism. This is due to the residual CPA part around $E=0$ that is never totally empty. We can also see that at $N_e \approx 0.4$ the maximum of the magnetisation is not at large $U$: $m$ increases at around $U\approx 8$ to reach its peak at $U\approx 2$.
Keeping the filling constant 
($N_e=0.4$ and $N_e=0.7$ respectively), while varying $\tau$ give the plots of 
Figs.\ref{pict:mUNe04} and \ref{pict:mUNe07}.
The magnetisation is interpolated between its VCA value at $\tau=0$ and the CPA value at $\tau=\infty$. At 
$\tau =0$ and for the fillings considered, the system is fully polarised at large enough $U$. The magnetisation has already dropped to zero for any $U$ at $\tau=10$ for $N_e=0.4$, and at $\tau=50$ for $N_e=0.7$. The increase in $m$ before the drop in magnetisation is particularly acute for $N_e=0.4$ and $\tau=2.0$, and is an artifact of the model caused by 
the position of the chemical potential in the VCA and CPA sub-bands. 
Fig.\ref{fig:dostau2} shows the respective positions of the CPA and the VCA sub-bands and the chemical potential 
for $N_e=0.4$, compared with $N_e=0.7$ at $\tau=2.0$.
At large $U$, the density of states of the up-spins (the minority spins) exhibits three well separated sub-bands for both fillings $N_e=0.4$ and $0.7$. The lower CPA sub-band is completely full. As $U$ decreases, the VCA band gets closer to the lower CPA sub-band, in both the up- and down-spin DOS. At first, only the down-spin bands are affected, meaning that the chemical potential travels down the energy scale while the proportion of up-spins and down-spins remains more or less constant. This is the situation illustrated in the two uppermost plots of Fig.\ref{fig:dostau2}. When $U$ is further decreased, the chemical potential will eventually reach the edge of the CPA lower sub-band. At a filling of $N_e=0.4$, this is happening at a fairly large $U$, $U\approx 8.0$, so that the VCA band is still well separated (plot in the middle of the left-hand side, Fig.\ref{fig:dostau2}). $N{\up}$ will thus decrease until the VCA band merges with the CPA band (bottom plot on the left, Fig.\ref{fig:dostau2}). For a higher filling, the two sub-bands merge before or at the same time as the chemical potential gets into the CPA band. The density of states at the chemical potential thus rapidly increases for the up-spins, and the magnetisation decreases accordingly as the up-spin band matches the down-spin band.\\
\noindent The increase in $m$ at intermediate $U$, though also occurring 
at other small fillings, becomes dramatic around $N_e\approx 0.4$ and $\tau\approx 2.0$. It is at this order of magnitude of $\tau$ that the system, at $N_e\approx 0.4$ , exhibits a balanced CPA-VCA behaviour. At much smaller $\tau$, the relative weight of the CPA band becomes very small compared to the VCA part. The chemical potential falls within the CPA sub-band of the up-spin over a large range of $U$ before the up-spin VCA band reaches it. At large $\tau$ the system becomes more CPA in character with a small VCA part responsible for the low magnetisation. 

\section{\label{sec:conclusion}Conclusion}
A new treatment of the Hubbard model is proposed that combines the physical 
pictures and respective advantages of the CPA and the VCA. It is based on the fact that the CPA on the Hubbard model considers the scatterers as fixed at their positions at $t=0$, whereas the VCA performs a time-average of the system.
The one-particle Green's function in the so-called $\tau$-CPA is built from the one-particle time Green's functions of the CPA and VCA. The CPA Green's function is applied up to a time $\tau$, thereafter the VCA Green's function, multiplied by a factor for continuity, is used. This time $\tau$ is taken to be of the order of magnitude of the time the spins travel between sites, that is $\tau$ has the value of the inverse of the band width.  The resulting Green's function, which must be first smoothed out and normalised, interpolates between the VCA propagator at $\tau=0$ and the CPA propagator at $\tau=\infty$. 
The density of states at finite $\tau$ exhibits both the CPA sub-bands (or one CPA band, if $U$ is small) and a VCA band. The bands are weighted according to the value of $\tau$. The CPA sub-bands are fixed at their positions at $E=0$ and $E=U$, while the VCA band, centred at $E=cU$, moves along with the concentration $c$. Equilibrium conditions with a polarised ground state are 
therefore found that are more stringent than the Stoner criterion for finite $\tau$.  \\
Artifacts of the $\tau$-CPA mixture include some non-regular features in the density of states for a range of parameters (intermediate $U$ and concentration), and peculiar behaviour of the $m$ versus $U$ curves at large $\tau$ and small concentration. The $\tau$-CPA is however a successful method for the Hubbard model that is based on the better physical picture given by the combination 
of the CPA and VCA. It allows the treatment of the scattering by the CPA 
while also predicting ordered ground state for large enough $U$. 
In a next publication, it will be shown that the $\tau$-CPA can also be 
used to calculate two-particle Green's functions. In particular the dynamical susceptibility in the Hubbard model can be evaluated using this method. 

\begin{acknowledgments}
AU acknowledges the support of the
Swiss National Science Foundation, the Berrow scholarship trust of Lincoln College and the ORS award scheme.
\end{acknowledgments}

\bibliographystyle{apsrev1} 
\bibliography{biblio1}

\end{document}